\documentclass[pdftex,twocolumn]{aastex6}

\usepackage{apjfonts,graphicx,graphics}

\usepackage{graphicx}
\usepackage{natbib}
\usepackage{mathptmx}
\usepackage{amsmath}
\usepackage{wasysym}

\usepackage{times}
\usepackage{epsfig}
\usepackage{ulem}

\usepackage{hyperref}
\hypersetup{tex4ht}
\usepackage{amsmath}	% Advanced maths commands
\usepackage{amssymb}	% Extra maths symbols
\usepackage[utf8]{inputenc}
\usepackage{graphics,epsfig,ulem,times}	% Makes sure all graphics works

\usepackage{tikz}				%tikz plotting
\usetikzlibrary{positioning}
\usetikzlibrary{external}
\usepackage{array}
\newcolumntype{C}[1]{>{\centering\arraybackslash}p{#1}}
\newcolumntype{?}{!{\vrule width 2.5\arrayrulewidth}}
\usepackage{multirow, makecell}
\begin{document}

% formed by enhanced cooling behind rising radio bubbles
% supported at low velocities beneath rising radio bubbles in PKS\,0745
%
%[ALMA observations of the Phoenix cluster]
%\altaffilmark{1}$^{*}$
\title{Biased tracer reconstruction with halo mass information}
\author{Yu Liu$^{1}$, Yu Yu$^{1}$, Baojiu Li$^{2}$}
\affil{$^1$Department of Astronomy, School of Physics and Astronomy,
Shanghai Jiao Tong University, Shanghai, 200240, P. R. China; yuyu22@sjtu.edu.cn\\
    $^2$ Institute for Computational Cosmology, Department of Physics, Durham University, South Road, Durham DH1 3LE, UK
     }
%\altaffiltext{*}{\href{www.baidu.com}{201521160006@mail.bnu.edu.cn}}
\begin{abstract}
Plenty of crucial information about our Universe is encoded in the cosmic large-scale structure (LSS). However, the extractions of these information are usually hindered by the nonlinearities of the LSS, which can be largely alleviated by various techniques known as the reconstruction. In realistic applications, the efficiencies of these methods are always degraded by many limiting factors, a quite important one being the shot noise induced by the finite number density of biased matter tracers (i.e., luminous galaxies or dark matter halos) in observations. In this work, we explore the gains of biased tracer reconstruction achieved from halo mass information, which can suppress shot noise component and dramatically improves the cross-correlation between tracer field and dark matter. To this end, we first closely study the clustering biases and the stochasticity properties of halo fields with various number densities under different weighting schemes, i.e., the uniform, mass and optimal weightings. Then, we apply the biased tracer reconstruction method to these different weighted halo fields and investigate how linear bias and observational mass scatter affect the reconstruction performance. Our results demonstrate that halo masses are critical information for significantly improving the performance of biased tracer reconstruction, indicating a great application potential for substantially promoting the precision of cosmological measurements [especially for baryon acoustic oscillations (BAO)] in the ambitious on-going and future galaxy surveys.
\end{abstract}

\keywords{Cosmology: Large-scale structure of the Universe: Reconstruction: Baryon Acoustic Oscillations}

\section{Introduction}
\label{sec:intro}

By measuring galaxy distribution in the Universe, ambitious on-going and future galaxy surveys (e.g., 4MOST (\citealt{2016SPIE.9908E..1OD}), PFS (\citealt{2014PASJ...66R...1T}), DESI (\citealt{2013arXiv1308.0847L}), LSST (\citealt{2009arXiv0912.0201L}), WFIRST (\citealt{2018arXiv180403628D}) and Euclid (\citealt{2011arXiv1110.3193L}), etc.) will map cosmic large-scale structure (LSS) with high precision, which can further improve the accuracy of cosmological parameter inferences and greatly deepen the understanding of our Universe. However, the LSS has non-linearly evolved into highly non-Gaussian in the late Universe, which leads to the observed signatures deviating from the theoretical predictions by linear theory (i.e., non-linear effects), makes statistical information leak into higher-order statistics, and especially blurs some critical information encoded in the LSS of early Universe (e.g., BAO, primordial non-Gaussianities and etc.), etc. These defects can induce undesired systematics in the analysis of observables, and thus limit the constraining power of cosmic probes on cosmological parameters and various promising candidates of new physics (e.g., dark energy models, modified gravity theories and inflation models, etc.).

For reducing nonlinearities in two-point statistics, Gaussianization methods (e.g., \citealt{1992MNRAS.254..315W}; \citealt{2009ApJ...698L..90N}) to some degree can increase information contents and alleviates mode coupling (\citealt{2011ApJ...731..116N}) by Gaussianizing the one-point probability function of the density field. Actually, a large part of the nonlinearities are caused by the large-scale bulk flows (\citealt{2007ApJ...664..660E}), which can move galaxies on average by approximately 10 $h^{-1}$Mpc from their initial locations (\citealt{2010ApJ...715L.185P}; \citealt{2014MNRAS.445.3152B}). Thus, these local transformation methods cannot genuinely convert the matter distribution back to its earlier stage with basically no improvement in the correlation between the final non-linear field and its initial condition (\citealt{2013MNRAS.436..759H}). Several different strategies have been proposed to tackle this problem, including the forward modeling (e.g., \citealt{2013ApJ...772...63W}; \citealt{2013MNRAS.432..894J}; \citealt{2013MNRAS.429L..84K}; \citealt{2017JCAP...12..009S}; \citealt{2018JCAP...07..043F}; \citealt{2018JCAP...10..028M}; \citealt{2019JCAP...11..023M}; \citealt{2019A&A...625A..64J}, etc.), backward reconstruction (hereafter reconstruction, e.g., \citealt{2002Natur.417..260F}; \citealt{2003MNRAS.346..501B}; \citealt{2007ApJ...664..675E}; \citealt{2012JCAP...10..006T}; \citealt{2015MNRAS.453..456B}; \citealt{2015PhRvD..92l3522S}; \citealt{2017PhRvD..96l3502Z}; \citealt{2017PhRvD..96b3505S}; \citealt{2017JCAP...09..012O}; \citealt{2018MNRAS.478.1866H}; \citealt{2018PhRvD..97b3505S}; \citealt{2019MNRAS.484.3818S}, etc.) and machine learning (e.g., \citealt{2020arXiv200210218M}), which can reproduce the initial density field at different levels.

In particular, the reconstruction techniques can directly reverse the bulk motions by estimating the displacement field based on the observed data, and have been commonly applied in BAO measurements and also have inspired many other cosmological applications [e.g., redshift space distortions (RSDs) (\citealt{2018PhRvD..97d3502Z}; \citealt{2019arXiv191203392W}), velocity reconstruction (\citealt{2019ApJ...887..265Y}), etc.]. Based on Zel'dovich approximation, \citealt{2007ApJ...664..675E} proposed a standard reconstruction algorithm for improving BAO measurement accuracy. This technique has been tested with simulated data (e.g., \citealt{2008ApJ...686...13S}; \citealt{2010ApJ...720.1650S}; \citealt{2011ApJ...734...94M}; \citealt{2014MNRAS.445.3152B}; \citealt{2015PhRvD..92h3523A}), and been theoretically studied (e.g., \citealt{2009PhRvD..80l3501N}; \citealt{2009PhRvD..79f3523P}; \citealt{2015MNRAS.450.3822W}; \citealt{2016MNRAS.460.2453S}; \citealt{2017PhRvD..96d3513H}; \citealt{2019JCAP...09..017C}; \citealt{2020PhRvD.101d3510H}), and also been extensively applied to observation data analysis (e.g., \citealt{2012MNRAS.427.2132P}; \citealt{2012MNRAS.427.2146X}; \citealt{2013MNRAS.431.2834X}; \citealt{2014MNRAS.441...24A}; \citealt{2014MNRAS.441.3524K}; \citealt{2015MNRAS.449..835R}; \citealt{2016MNRAS.455.3230B}; \citealt{2017MNRAS.464.3409B}; \citealt{2017MNRAS.464.4807H}). In recent years, several improved new reconstruction algorithms [e.g., the isobaric reconstruction technique (\citealt{2017PhRvD..96l3502Z}), the iterative reconstruction technique (\citealt{2017PhRvD..96b3505S}; \citealt{2018MNRAS.478.1866H}), the multigrid relaxation method (\citealt{2018PhRvD..97b3505S}), the extended fast action minimisation method (\citealt{2019MNRAS.484.3818S}) and the fast semi-discrete optimal transport algorithm (\citealt{2020arXiv201209074L})] were proposed and tested not only on matter field (e.g., \citealt{2017ApJ...841L..29W}; \citealt{2017MNRAS.469.1968P}) but also on more realistic halo/galaxy field (e.g., \citealt{2017ApJ...847..110Y}; \citealt{2019ApJ...870..116W}; \citealt{2019MNRAS.483.5267B}; \citealt{2019MNRAS.482.5685H}; \citealt{2020arXiv201010456S}), showing that they can substantially bring back initial information and recover the linear BAO signal. %, which is beyond the performance of standard reconstruction.

The observed halos/galaxies are discrete and biased tracers of the underlying dark matter field. When applied to biased tracer field, these reconstruction techniques will be largely limited by additional complications (e.g., halo/galaxy bias and shot noise, etc.), which can physically or numerically affect the reconstruction process, leading to a worse estimation of the displacement field compared to the case of the matter field, and then consequently induce larger errors in the recovery of the initial matter field and degrade their powers in cosmological applications especially for the reconstruction of BAO wiggles (\citealt{2017ApJ...847..110Y}; \citealt{2018MNRAS.479.1021D}; \citealt{2019ApJ...870..116W}; \citealt{2019MNRAS.483.5267B}). Thus, in observations, these practical issues should be addressed or taken into account for further improving the reconstruction performance.

Recently, the effects of linear bias on BAO isobaric reconstruction were theoretically investigated and modeled (\citealt{2019ApJ...870..116W}). Also, \citealt{2019MNRAS.483.5267B} developed a biased tracer reconstruction technique by extending the method proposed in \citealt{2018PhRvD..97b3505S} to include bias scheme up to the quadratic order. Based on mass conservation, this method transforms the reconstruction problem into solving a Monge-Amp\`ere-type equation, which can be numerically solved by the multigrid relaxation method. It was demonstrated that this biased tracer reconstruction method can help further substantially improve the recovery of initial density and linear BAO wiggles from the biased tracer field, by performing debiasing in the process of reconstruction.

With these major progresses in biased tracer reconstruction, the difficulties caused by halo/galaxy bias have been alleviated now, while the shot noise (due to the limited number density of tracers) is still a significant limiting factor in realistic scenarios. For mitigating this critical problem, additional information should in principle be taken into account in constructing tracer field. The most natural idea could be the considerations of halo masses or galaxy luminosities (or stellar masses), which can be directly measured or indirectly inferred in observations (e.g., \citealt{2004MNRAS.353..189V}; \citealt{2005ApJ...633..791Z}; \citealt{2005MNRAS.356.1293Y}; \citealt{2018MNRAS.481.5470X}; \citealt{2018ARA&A..56..435W} and Refs. therein). Interestingly, \citealt{2009PhRvL.103i1303S} found that weighting halos by their masses indeed can suppress halo field's shot noise component and can tighten the correlation between halo field and underlying dark matter (i.e., reducing the stochasticity), dramatically improving upon the commonly used uniform weighting scheme (i.e., weighting the halos/galaxies uniformly). Afterwards, \citealt{2010PhRvD..82d3515H} developed an optimal mass-dependent halo weighting technique. Comparing to the previous mass weighting, this optimal weighting scheme can further suppress shot noise and minimizes the stochasticity, which was also confirmed by \citealt{2011MNRAS.412..995C} by using a different methodology. Then, these findings triggered several specific applications [e.g., constraints on primordial non-Gaussianity (\citealt{2011PhRvD..84h3509H}) and growth rate of structure formation (\citealt{2012PhRvD..86j3513H}), etc.] and many other cosmological investigations (e.g., \citealt{2013PhRvD..88h3507B}; \citealt{2015MNRAS.446..793J}; \citealt{2016MNRAS.457.2968S}; \citealt{2017PhRvD..96h3528G}; \citealt{2019PhRvD.100d3514S}, etc.).

Motivated by these progresses, in this work we aim to extend the study of \citealt{2019MNRAS.483.5267B} with the goal of further improving the bias tracer reconstruction performance by including information of halo masses. The biased tracer reconstruction method adopted in this work uses mass conservation to relate the density fields in Lagrangian and Eulerian spaces. Naturally, we expect a great improvement in the recovery of the initial matter field when the method is performed on a halo field with more information of underlying matter distribution, which can be obtained from the information of halo masses. To corroborate this idea, in our work, we systematically perform careful analyses on various halo fields under different halo weighting schemes, i.e., the uniform, mass and optimal weightings (the latter two weightings are both related to halo masses). Our results suggest that halo masses undoubtedly are important information for highly improving the performance of biased tracer reconstruction, thus demonstrating enormous application potential for significantly improving the power of extracting cosmological information from the ambitious current and future galaxy surveys.

This paper is organized as follows. We review briefly the biased tracer reconstruction method in Section \ref{sec:method}. In Section \ref{sec:data}, we introduce in detail the data used in this work. The Section \ref{sec:bias and stochasticity} formulates the definitions of halo clustering bias and stochasticity. The results of our investigations are presented in Section \ref{sec:results}. Finally, in Section \ref{sec:conclusions}, we summarize our work, give discussions and draw our conclusion.

\begin{table*}
\center
\begin{tabular}{p{1.0cm}p{4.0cm}<{\centering}p{3.9cm}<{\centering}p{3.9cm}<{\centering}p{3.3cm}<{\centering}}
\hline\hline
Redshift & Mass range $[h^{-1}\mathrm{M}_\odot]$ & Number density $[(h^{-1}\mathrm{Mpc})^{-3}]$ & Linear bias parameter & The galaxy survey\\
\hline\hline
%\midrule
$z=0$ & $M_{h}$ $\in$ $[2.15\times 10^{10},2.11\times 10^{15}]$ & $\bar{n}_h=2.77\times 10^{-2}$ & $b_U=0.68$, $b_M=1.52$, $b_O=1.10$ & LSST (\citealt{2009arXiv0912.0201L}); DESI BGS (\citealt{2016arXiv161100036D})\\
$z=0$ & $M_{h}$ $\in$ $[1.84\times 10^{12},2.11\times 10^{15}]$  & $\bar{n}_h=2.77\times 10^{-3}$ & $b_U=0.94$, $b_M=1.75$, $b_O=1.35$ & DESI ELG; SPHEREx (\citealt{2014arXiv1412.4872D})\\
$z=0$ & $M_{h}$ $\in$ $[2.10\times 10^{13},2.11\times 10^{15}]$  & $\bar{n}_h=2.77\times 10^{-4}$ & $b_U=1.55$, $b_M=2.30$, $b_O=1.95$ & PFS (\citealt{2014PASJ...66R...1T}); BOSS CMASS (\citealt{2013AJ....145...10D}); Euclid (\citealt{2011arXiv1110.3193L})\\
\hline
\end{tabular}\\
\caption{Various halo samples used in the reconstruction performance tests. Note that our analyses are based on the halo samples at $z=0$, which is inconsistent with the redshifts of targets of these surveys. Since the primary goal of this work is to investigate the gains of bias tracer reconstruction achieved from halo mass information, this inconsistency won't affect our conclusions.
}
\label{tab:info}
\end{table*}

\section{The biased Tracer Reconstruction Method}
\label{sec:method}

With the assumptions of mass conservation and uniform initial matter distribution\footnote{Based on the same basic assumptions, the reconstruction problem can also be treated as an optimized mass transportation problem which can be solved by using the Monge-Amp\`ere-Kantorovich (MAK) method (see \citealt{2002Natur.417..260F}; \citealt{2003MNRAS.346..501B}; \citealt{2003A&A...406..393M}; \citealt{2006MNRAS.365..939M}; \citealt{2008PhyD..237.2145M}). For a sample with $N$ points, the fully-discrete combinatorial algorithms of MAK reconstruction have a complexity of $N^3$, which thus hampers their applications to big data samples in LSS study. Very recently, \citealt{2020arXiv201209074L} reported a new semi-discrete algorithm with an empirical complexity of $N{\rm log}N$, which makes it significantly more efficient than the previous combinatorial ones.},
\begin{equation}\label{eq:1}
\rho({\bf x}){\rm d}^3{\bf x} = \rho({\bf q}){\rm d}^3{\bf q} \approx \bar{\rho}{\rm d}^3{\bf q},
\end{equation}
there is an unique mapping from the initial Lagrangian coordinates $\textbf{q}$ to the final Eulerian coordinates $\textbf{x}$ before shell crossing of structure formation (cf. \citealt{2002Natur.417..260F}; \citealt{2003MNRAS.346..501B} and Refs. therein),
\begin{equation}\label{eq:2}
\det\left(\frac{\partial q^i}{\partial x^j}\right) = \det\left[\nabla^{i}\nabla_{j}\Theta(\bf{x})\right] = \frac{\rho(\bf{x})}{\bar{\rho}} \equiv 1 +\delta_m\left({\bf x}\right),
\end{equation}
where $\det\left(\frac{\partial q^i}{\partial x^j}\right)$ is the Jacobian determinant of mapping from  $\textbf{q}$ to $\textbf{x}$, $\delta_m\left({\bf x}\right)$ is the dark matter overdensity (hereafter, we also refer to it as {\it dark matter field}) and the displacement potential $\Theta$ is defined by $\mathbf{q} = \nabla_{\mathbf{x}}\Theta(\mathbf{x})$. Actually, with the nonlinear evolution of LSS the shell crossing is inevitable on small scales in the late Universe, which makes the mapping no longer unique and thus leads to meaningless reconstruction results below the shell-crossing scale. Nevertheless, the assumption of no shell crossing in the reconstruction process is indeed valid on relatively larger scales, which guarantees that the algorithm can work well to recover the large-scale initial information (cf. \citealt{2018PhRvD..97b3505S}; \citealt{2019MNRAS.483.5267B}). This is particularly useful for its cosmological applications (e.g., the BAO and primordial non-Gaussianity measurements, etc.), because the large-scale information is more cosmologically relevant.

By expanding $\det\left[\nabla^{i}\nabla_{j}\Theta({\bf x})\right]$, equation (\ref{eq:2}) can be rewritten as
\begin{eqnarray}\label{eq:3}
&&\frac{1}{6}\left(\nabla^2\Theta\right)^3 - \frac{1}{2}\nabla^i\nabla_j\Theta\nabla^j\nabla_i\Theta\nabla^2\Theta\nonumber\\
&&~~~~~~~~~~~~~~ + \frac{1}{3}\nabla^i\nabla_j\Theta\nabla^j\nabla_k\Theta\nabla^k\nabla_i\Theta\ =\ 1 +\delta_m\left({\bf x}\right),
\end{eqnarray}
where the Einstein summation convention is used. Equation (\ref{eq:3}) is a non-linear elliptical partial differential equation (PDE), which can be numerically solved. For the convenience of numerical implementation, $\nabla^i\nabla_j\Theta$ should be split into diagonal and traceless parts by defining barred derivatives (see \citealt{2018PhRvD..97b3505S}; \citealt{2019MNRAS.483.5267B} for more discussions),
\begin{equation}\label{eq:4}
	\nabla^i\nabla_j\Theta \equiv \dfrac{1}{3}\delta^i_{j}\nabla^2\Theta + \bar{\nabla}^i\bar{\nabla}_j\Theta,
\end{equation}
where $\delta^i_{j}$ is the Kronecker delta.
Thus, the equation (\ref{eq:3}) is further rewritten as
\begin{eqnarray}\label{eq:5}
(\nabla^2\Theta)^3-\frac{9}{2}\bar{\nabla}^i\bar{\nabla}_j\Theta\bar{\nabla}^j\bar{\nabla}_i\Theta\nabla^2\Theta~~~~~~~~~~~~~~~~~~~~~~~~~~~ && \nonumber\\
+ 9\bar{\nabla}^i\bar{\nabla}_j\Theta\bar{\nabla}^j\bar{\nabla}_k\Theta\bar{\nabla}^k\bar{\nabla}_i\Theta -27[1+\delta_m(\bf{x})] &=& 0,~~~
\end{eqnarray}
which is called the {\it reconstruction equation}.

However, the dark matter field $\delta_m({\bf x})$ is not actually observable. In galaxy surveys, we measure the biased tracer field, i.e., $\delta_{\rm tracer}({\bf x})\equiv n_{\rm tracer}({\bf x})/\bar{n}_{\rm tracer}-1$, where $n_{\rm tracer}({\bf x})$ is the tracer number density and $\bar{n}_{\rm tracer}$ is its mean value. The connection between them can be described by a series of bias parameters. Here, we only consider the bias parameters up to second order, and their relation is expressed as
\begin{equation}\label{eq:6}
	\delta({\bf x})_{\rm tracer} = b_1\delta_m({\bf x}) + \dfrac{b_2}{2}\delta_m^2({\bf x}) + \gamma_2\mathcal{G}_2,
\end{equation}
where $b_1$, $b_2$, $\gamma_2$ are the linear, quadratic, non-local bias parameters respectively and the non-local bias term  $\mathcal{G}_2$ is related to the velocity potential  $\Phi_v$ (see \citealt{2012PhRvD..85h3509C}; \citealt{2019MNRAS.483.5267B} for more details).
Replacing $\delta_m({\bf x})$ with $\delta_{\rm tracer}({\bf x})$ in equation (\ref{eq:5}), then we obtain a more general version of reconstruction equation,
\begin{eqnarray}\label{eq:7}
(\nabla^2\Theta)^3 - \dfrac{9}{2}\bar{\nabla}^i\bar{\nabla}_j\Theta\bar{\nabla}^j\bar{\nabla}_i\Theta\nabla^2\Theta + 9\bar{\nabla}^i\bar{\nabla}_j\Theta\bar{\nabla}^j\bar{\nabla}_k\Theta\bar{\nabla}^k\bar{\nabla}_i\Theta~~~ && \nonumber\\
- 27\left[1+\dfrac{\delta_{\rm tracer}({\bf x})}{b_1} - \dfrac{b_2}{2b_1^3}\delta_{\rm tracer}^2({\bf x}) - \dfrac{\gamma_2}{b_1}\left(\nabla^i\nabla_j\Theta\nabla^j\nabla_i\Theta \right.\right. \nonumber\\ \left.\left. + 4\nabla^2\Theta - \left(\nabla^2\Theta\right)^2 - 6\right)\right] = 0. ~~~~~~~~~~~~~~~~~~~~~~~~~~~~~~~~~~~~~~~~
\end{eqnarray}
If $b_1=1$ and $\gamma_2=b_2=0$, the equation (\ref{eq:7}) will be reduced to equation (\ref{eq:5}). From now on, we call equation (\ref{eq:7}) the {\it biased tracer reconstruction equation}.

Following the methodology formulated in \citealt{2018PhRvD..97b3505S} and \citealt{2019MNRAS.483.5267B}, equation (\ref{eq:7}) can be iteratively solved for $\Theta(\bf x)$ and $\nabla_{\mathbf{x}}\Theta(\mathbf{x})$ on discrete mesh cells by using multigrid Gauss–Seidel technique. The algorithm is implemented by modifying the code of \textsc{ecosmog} (\citealt{2012JCAP...01..051L}; \citealt{2013JCAP...11..012L}), which is based on the publicly available N-body code \textsc{ramses} (\citealt{2002A&A...385..337T}). Then, the displacement field is given by ${\bf \Psi}({\bf q}) = \textbf{x} - \textbf{q}$, where $\mathbf{q}(\mathbf{x}) = \nabla_{\mathbf{x}}\Theta(\mathbf{x})$. Based on the linear Lagrangian perturbation theory (LPT), the reconstructed density field is finally obtained by the negative divergence of displacement field with respect to $\textbf{q}$
\begin{equation}
	\delta_r = -\nabla_{\mathbf{q}} \cdot\Psi\left({\bf q}\right),
	\label{eq:8}
\end{equation}
which indicates that only the curl-less “E-mode” component of $\Psi\left({\bf q}\right)$ is used for initial information recovery (\citealt{2017PhRvD..95d3501Y}). Here, the divergence is implemented using the publicly available \textsc{dtfe} (Delaunay Tessellation Field Estimator) code (\citealt{2000A&A...363L..29S}; \citealt{2011ascl.soft05003C}), which is based on Delaunay tessellation.

\section{Data}
\label{sec:data}

\subsection{$N$-body simulation and halo samples}
\label{sec:simulation and halos}

In this work, we adopt a cosmological $N$-body simulation realized using publicly-available code \textsc{cubep$^3$m} (\citealt{2013MNRAS.436..540H}). This simulation is initialized at redshift $z=127$ and evolves $2048^3$ CDM particles with mass resolution of $2.15\times10^{9}$ $h^{-1}M_{\odot}$ in a periodic cubic box of width $600$ $h^{-1}$Mpc. Here, a Particle-Mesh process with $4096^3$ grids is used for gravitational force calculation. For increasing force resolution below the mesh scale, a Particle-Particle algorithm is also involved.

The \textsc{cubep$^3$m}’s own on-the-fly spherical overdensity (SO) halo finder is employed to identify CDM halos, the masses of which are resolved down to $2.15\times10^{10}$ $h^{-1}M_{\odot}$ with a minimum of 10 CDM particles per halo. We intend to investigate the dependence of reconstruction performances on different halo number densities (i.e., different shot noise levels). To this end, three halo samples with number densities of $2.77\times10^{-2}$ ($h^{-1}$Mpc)$^{-3}$, $2.77\times10^{-3}$ ($h^{-1}$Mpc)$^{-3}$ and $2.77\times10^{-4}$ ($h^{-1}$Mpc)$^{-3}$ are constructed by discarding the halos with masses below the mass cutoffs of $M_{\rm min} \simeq 2.15\times10^{10}$ $h^{-1}M_{\odot}$, $M_{\rm min} \simeq 1.84\times10^{12}$ $h^{-1}M_{\odot}$ and $M_{\rm min} \simeq 2.10\times10^{13}$ $h^{-1}M_{\odot}$ respectively. The number densities of our halo samples are roughly comparable to those of targets of various galaxy surveys listed in table \ref{tab:info}. These halo samples were also used in \citealt{2017ApJ...847..110Y}, where the isobaric reconstruction method was adopted for non-linear halo reconstruction without considering halo mass and bias etc. Note that throughout this paper, subhalos are excluded from the analyses.

\subsection{Different weighted halo fields}
\label{sec:halo fields}

If one halo sample is divided into $N$ mass bins and halos in the $i$-th bin are assigned the same weight $w(M_i)$, the weighted halo field $\delta_w({\bf x})$ in configuration space can be written as
\begin{equation}\label{eq:9}
\delta_w({\bf x})=\dfrac{\Sigma_iN_iw(M_i)\delta_i({\bf x})}{\Sigma_iN_iw(M_i)},
\end{equation}
where $N_i$ is the number of halos in the $i$-th bin and the uniformly weighted halo overdensity $\delta_i({\bf x})$ of the $i$-th bin is defined as
\begin{equation}\label{eq:10}
\delta_i({\bf x})\equiv\dfrac{n_i({\bf x})}{\bar{n_i}}-1,
\end{equation}
here $n_i({\bf x})$ and $\bar{n_i}$ are the local and mean halo number density of the $i$-th bin respectively. We denote by $w(M_i)=1,M_i,b(M_i)$ and $W_{opt}(M_i)$ the uniform, mass, bias\footnote{Various previous works (e.g., \citealt{2009PhRvL.103i1303S}; \citealt{2010PhRvD..82d3515H}; \citealt{2011MNRAS.412..995C}) found that the bias weighting only can lead to marginal or even basically no improvements relative to the uniform weighting for suppressing halo stochasticity, which is also confirmed in our check. This is because that the bias weighting is close to the uniform weighting which makes the implementation of local mass and momentum conservation inefficient for LSS evolution (\citealt{2009PhRvL.103i1303S}). Therefore, the investigations on the gains of bias tracer reconstruction achieved from the bias weighting scheme are omitted in this paper.} and optimal weighting schemes, respectively, for constructing the weighted halo field $\delta_w({\bf x})$, where $M_i$ is the average mass of halos in the $i$-th bin and the "optimal" can refer to the weighting scheme that minimizes the stochasticity of $\delta_w({\bf x})$ with respect to the underlying dark matter (\citealt{2010PhRvD..82d3515H}; \citealt{2011MNRAS.412..995C}). \citealt{2010PhRvD..82d3515H} states that the "optimal" weighting corresponds to a nontrivial eigenvector of the stochasticity matrix\footnote{In \citealt{2010PhRvD..82d3515H}, multiple mass bins with an equal number of halos are used for the derivation. The stochasticity matrix is defined as $C_{ij}\equiv\langle(\delta_i({\bf k})-b_i(k)\delta_m({\bf k}))(\delta_j({\bf k})-b_j(k)\delta_m({\bf k}))^*\rangle$ in Fourier space, where the bias of the $i$-th halo bin is determined by $b_i(k)=\dfrac{\langle\delta_i({\bf k})\delta_m({\bf k})^*\rangle}{\langle\vert\delta_m({\bf k})\vert^2\rangle}$.}, which yields the lowest eigenvalue. The eigenvector can be well fitted by a sample function with the form of $w(M) = M + M_0$, where $M_0$ is a free parameter.

\begin{figure*}
\begin{minipage}{\textwidth}
\center
\raisebox{0.0cm}{\includegraphics[width=0.93\columnwidth]{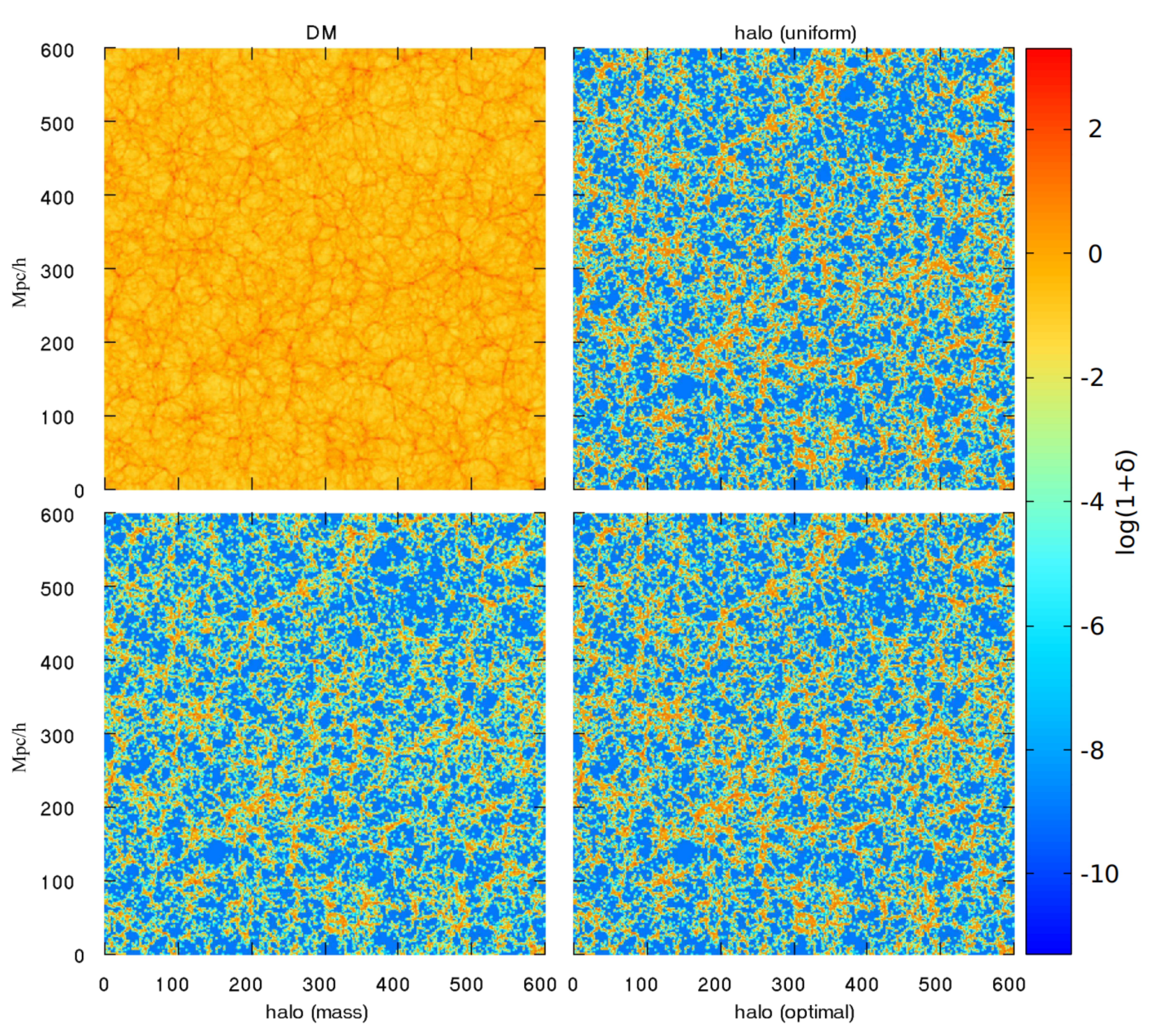}}
\caption{The two-dimensional slices of dark matter field (top-left) and different weighted halo fields with halo number density of $2.77\times10^{-2}$ $(h^{-1}{\rm Mpc})^{-3}$ before reconstruction, at $z = 0$. These fields are constructed by the CIC method. Each slice shows the same $600\times600$ $(h^{-1}{\rm Mpc})^2$ region with $5.86$ $h^{-1}{\rm Mpc}$ thickness of the simulation box.}
\label{fig:density_field_before_reconstruction}
\end{minipage}
\end{figure*}

Actually, except for the uniform weighting, increasing the number of halo bins can further suppress the halo stochasticity (\citealt{2010PhRvD..82d3515H}). Thus, in practical applications, each halo of mass $M$ should be assigned a weight $w(M)$ according to a smooth weighting function, corresponding to the number of bins approaching infinity (i.e., $N\rightarrow +\infty$). In our following performance tests, the smooth uniform, mass and optimal weighting functions are adopted with the forms of $w(M)=1,M, M + M_0$ respectively. Note that the last two weightings are both dependent on the information of halo masses and thus are also collectively called as mass-dependent weightings in this paper. In \citealt{2010PhRvD..82d3515H}, the $M_0$ was optimized iteratively by minimizing the $\delta_{w}({\bf x})$'s shot noise level, and they found the best value of $M_0$ depends on the cutoff mass $M_{min}$ and roughly satisfies the relation of $M \simeq 3M_{min}$ within their tested domain.

However, the simulation used in this work has higher mass resolution and resolves halos with lower masses. The halo finder algorithm applied here (the SO method) is also different from theirs (the friends-of-friends method). Considering these differences, the empirical relation $M_0 \simeq 3M_{min}$ may not hold in our cases, in particular for our halo sample with the highest number density (i.e., $2.77\times10^{-2}$ ($h^{-1}$Mpc)$^{-3}$). Moreover, in our study, we prefer to seek out the optimal halo weighting for maximazing the performance of bias tracer reconstruction, which not only depends on the $\delta_{w}({\bf x})$’s shot noise level but also on the bias of $\delta_{w}({\bf x})$ with respect to dark matter. Given all that, in our work the optimal $M_0$ is determined by maximizing the reconstruction performance (rather than minimizing the shot noise level of $\delta_{w}({\bf x})$, e.g., \citealt{2010PhRvD..82d3515H}). For the three halo samples with the number densities in descending order, we find the optimal $M_0$'s are approximately $70M_{min}\simeq1.5\times10^{12}$ $h^{-1}\mathrm{M}_\odot$, $6M_{min}\simeq1.1\times10^{13}$ $h^{-1}\mathrm{M}_\odot$, $3M_{min}\simeq6.3\times10^{13}$ $h^{-1}\mathrm{M}_\odot$, respectively.

\subsection{The mass assignment method}
\label{sec:mass assignment}

\citealt{2017ApJ...847..110Y} opted for the DTFE (i.e., Delaunay Tessellation Field Estimator) mass assignment (\citealt{2011arXiv1105.0370C}) to generate halo fields, for avoiding the instability of the isobaric reconstruction algorithm due to the sparseness of the halo samples. However, the DTFE scheme can induce excessive smoothing of the low-density regions of the halo field through a special window function of Delaunay tessellation, which inevitably erases some spatial distribution information of halos, thus weakening the reconstruction performance (\citealt{2017ApJ...847..110Y}; \citealt{2019MNRAS.483.5267B}). This halo sparseness issue can be tackled well by the biased tracer reconstruction technique as demonstrated in \citealt{2019MNRAS.483.5267B}. They found that the traditional cloud-in-cell (CIC) and triangular-shaped-cloud mass assignments can achieve similar reconstruction performance and present great improvements over the DTFE method. Given that the CIC method is more commonly used, in our work, the matter/halo fields are constructed via CIC interpolation of particles/halos onto cubical meshes with $512^3$ grids (cf. figure \ref{fig:density_field_before_reconstruction}), where the grid cell size ($\sim1.2$ $h^{-1}$Mpc) is sufficient for the convergence of the biased tracer reconstruction performance (\citealt{2019MNRAS.483.5267B}).

\section{bias and stochasticity}
\label{sec:bias and stochasticity}

\begin{figure*}
\begin{minipage}{\textwidth}
\raisebox{0.1cm}{\includegraphics[width=1\columnwidth]{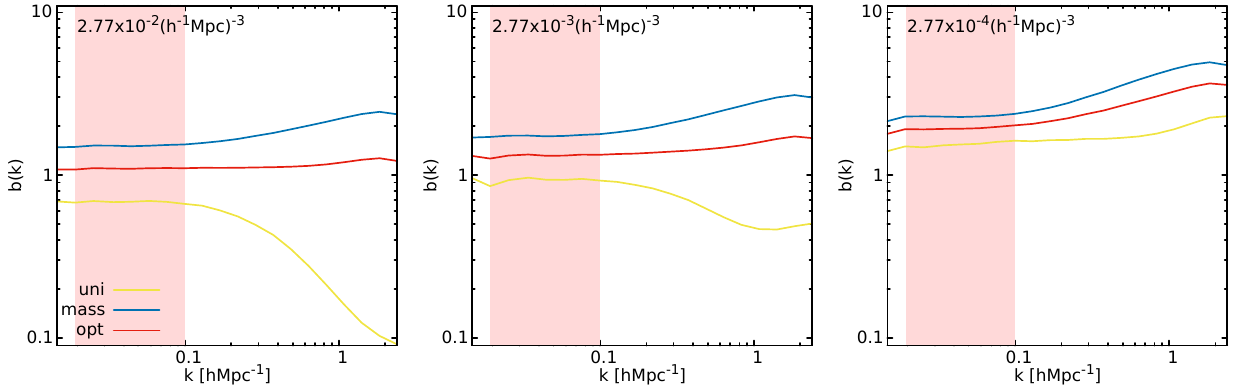}}
\caption{The clustering biases of different weighted halo fields. From left to right, the three panels correspond to different halo samples with number densities of $2.77\times10^{-2}$, $2.77\times10^{-3}$ and $2.77\times10^{-4}$ $(h^{-1}{\rm Mpc})^{-3}$, respectively (the same below). The yellow, blue and red curves correspond to the uniform, mass and optimal weighting schemes (the same below). The shaded regions indicate the scales of $0.025$ $h$Mpc$^{-1} \lesssim k \lesssim 0.099$ $h$Mpc$^{-1}$ for calculating the linear bias parameters.}
\label{fig:bias}
\end{minipage}
\end{figure*}

\begin{figure*}
\begin{minipage}{\textwidth}
\raisebox{0.1cm}{\includegraphics[width=1\columnwidth]{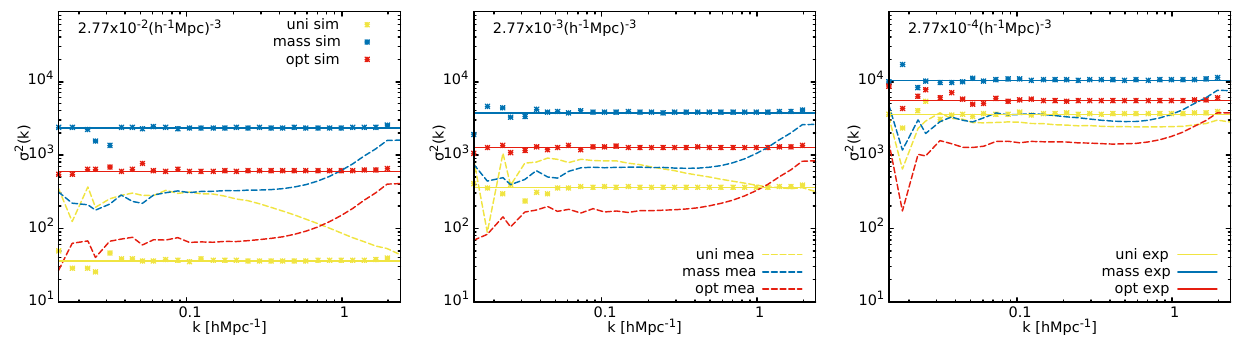}}
\caption{Various shot noises under different weighting schemes. The asterisk points, dashed lines and solid lines correspond to the simulated, measured and expected shot noises, respectively.}
\label{fig:shot_noise}
\end{minipage}
\end{figure*}

\begin{figure*}
\begin{minipage}{\textwidth}
\raisebox{0.1cm}{\includegraphics[width=1\columnwidth]{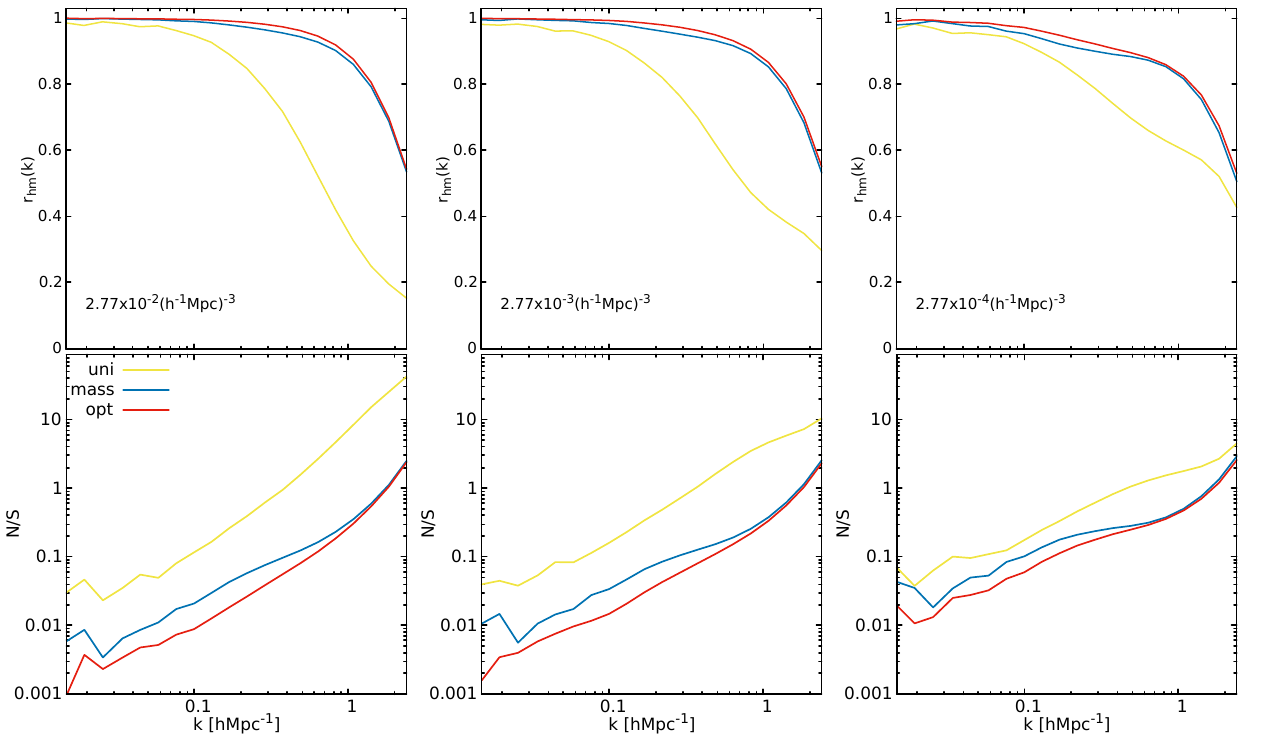}}
\caption{Top: The cross-correlation coefficients between different weighted halo fields and dark matter field, $r_{hm}(k)$. Bottom: The N/S ratios of different weighted halo power spectra, defined as $\frac{1-r_{hm}^2(k)}{r_{hm}^2(k)}$ (cf. equation \ref{eq:17}). The N/S ratios can be used to amplify the differences at large scales where the $r_{hm}$s' are very close to unity.}
\label{fig:stochasticity}
\end{minipage}
\end{figure*}

\begin{figure*}
\begin{minipage}{\textwidth}
\raisebox{0.1cm}{\includegraphics[width=1\columnwidth]{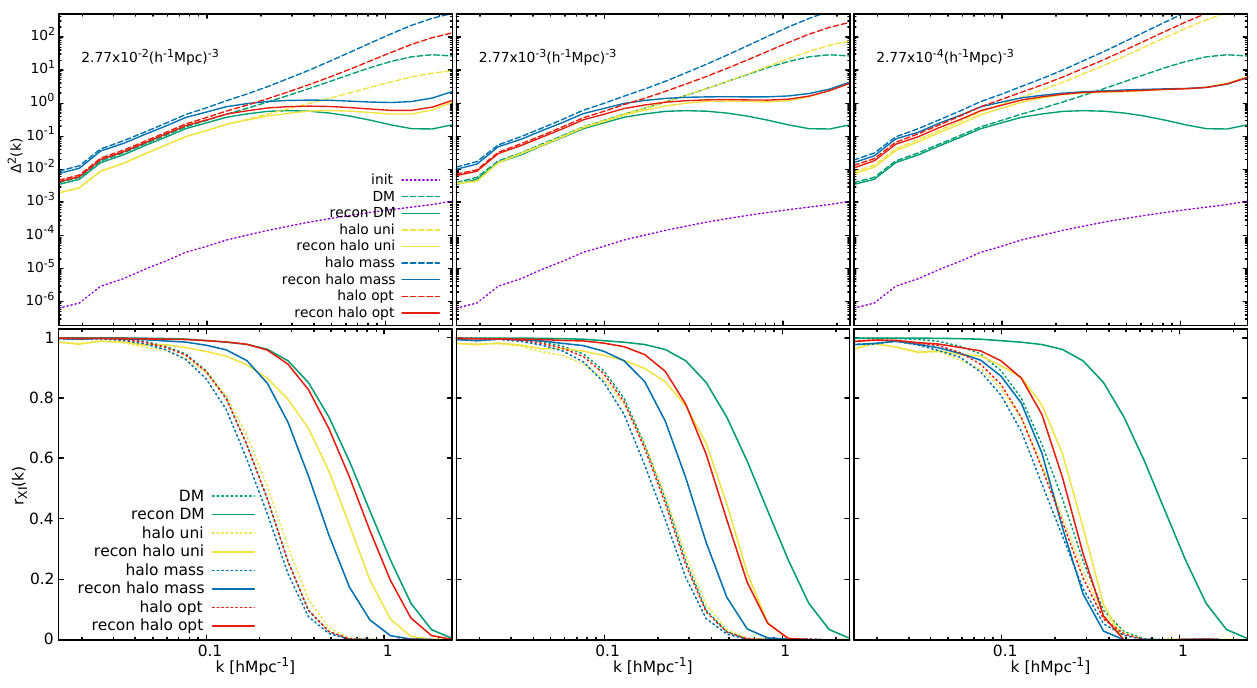}}
\caption{Top: The power spectra of different fields. The dashed lines and solid lines correspond to the fields before and after reconstruction, respectively. The green curves are for the dark matter field (the same below). For reference, we also plot the power spectrum of initial condition, which is presented by the dotted blue line. Bottom: The cross-correlation coefficients with initial condition for different fields of pre- and post-reconstruction. The cross-correlation coefficient between reconstructed dark matter field and initial condition (solid green line) is used to serve as an upper bound of the performance of biased tracer reconstruction. Note that here we perform the biased tracer reconstruction directly without preprocessing halo clustering bias.}
\label{fig:reconstruction_without_bias}
\end{minipage}
\end{figure*}

\begin{figure*}
\begin{minipage}{\textwidth}
\center
\raisebox{0.0cm}{\includegraphics[width=0.93\columnwidth]{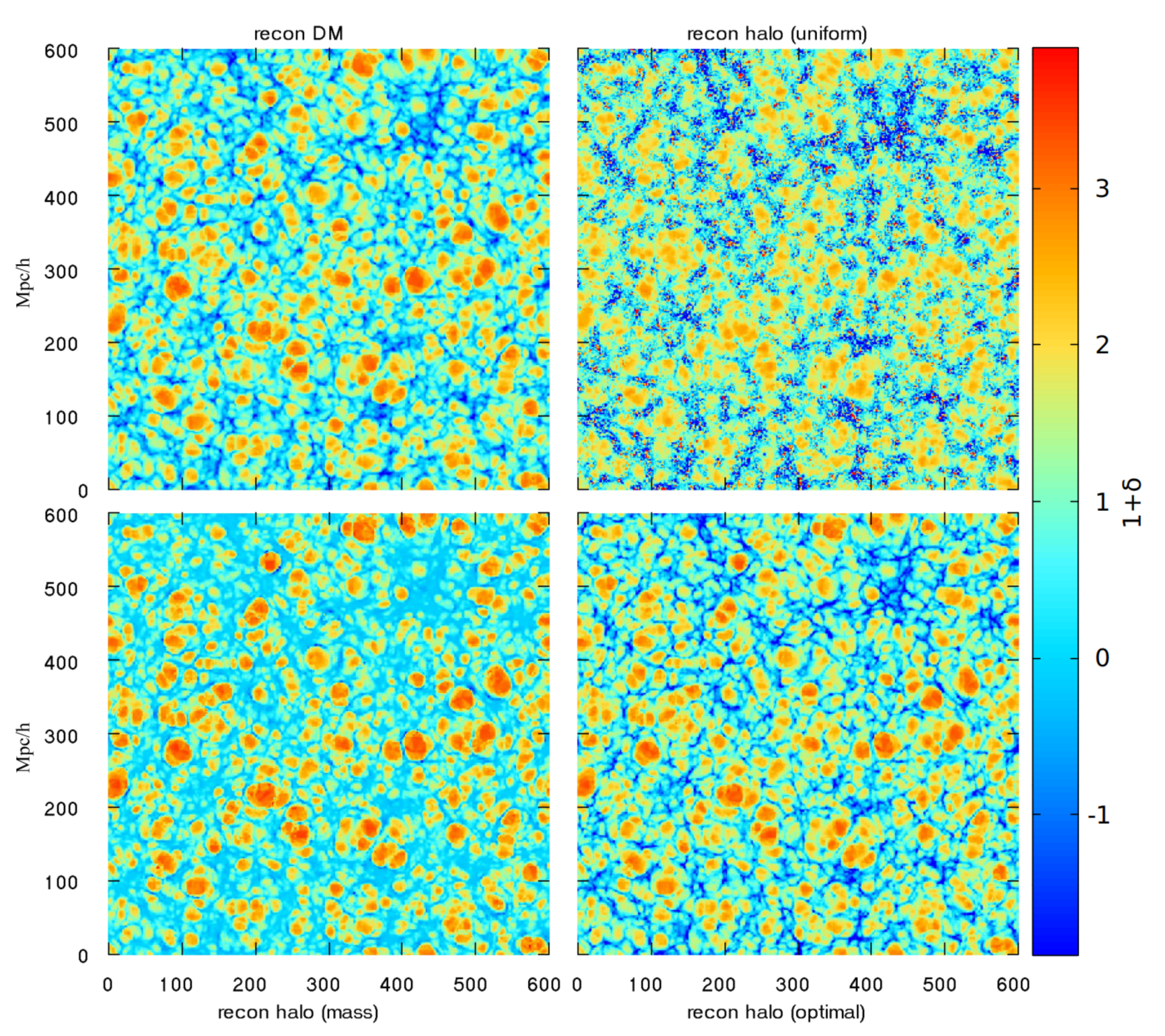}}
\caption{The two-dimensional slices of dark matter field (top-left) and different weighted halo fields with halo number density of $2.77\times10^{-2}$ $(h^{-1}{\rm Mpc})^{-3}$ after reconstruction. Each slice shows the same region (with the same projection depth) as the corresponding slice in figure \ref{fig:density_field_before_reconstruction}. Note that the linear biases are corrected in the halo cases.}
\label{fig:density_field_after_reconstruction}
\end{minipage}
\end{figure*}

\begin{figure*}
\begin{minipage}{\textwidth}
\raisebox{0.0cm}{\includegraphics[width=1\columnwidth]{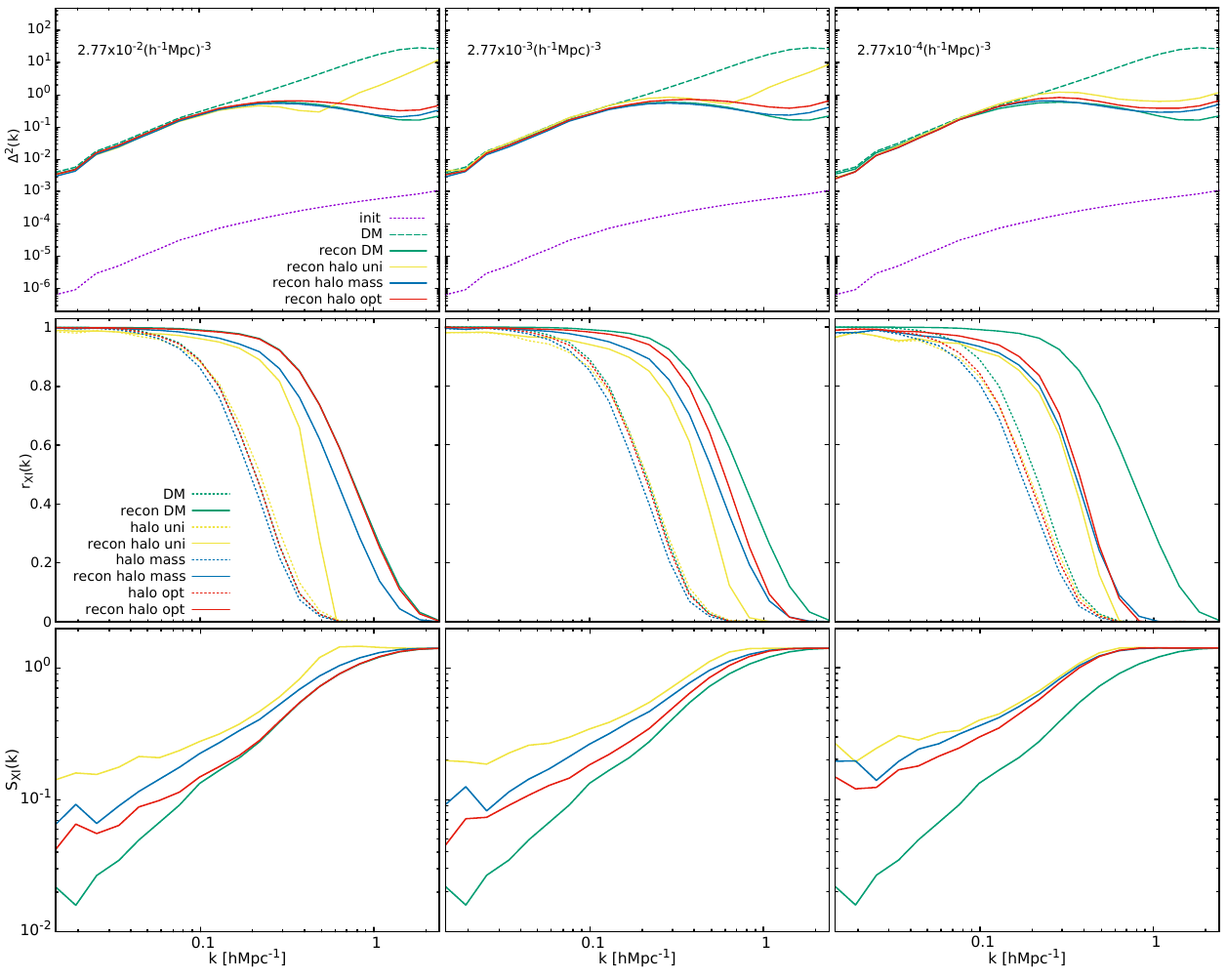}}
\caption{Top: The power spectra of different reconstructed fields (solid lines). For comparison, we also plot the power spectra of pre-reconstructed dark matter field (dashed green line) and initial condition (dotted blue line). Middle: The cross-correlation coefficients with initial condition for different fields of pre- and post-reconstruction, $r_{XI}(k)$. Bottom: The $S_{XI}(k)$. The $S_{XI}(k)$ is defined as $S_{XI}(k) \equiv \sqrt{2(1 - r_{XI}(k))}$, which can help amplify the differences at large scales where the $r_{XI}$s' are too close to unity. Here, we consider the linear clustering bias to perform the biased tracer reconstruction.}
\label{fig:reconstruction_with_bias}
\end{minipage}
\end{figure*}

Forming in the density peak regions, halos are merely biased and stochastic tracers of the underlying dark matter field. In Fourier space, the relation between halo overdensity $\delta_h({\bf k})$ and dark matter overdensity $\delta_m({\bf k})$ can be expressed as
\begin{equation}\label{eq:11}
\delta_h({\bf k}) = b(k)\delta_m({\bf k}) + \varepsilon({\bf k}),
\end{equation}
where the bias is defined as
\begin{equation}\label{eq:12}
b(k)=\dfrac{\langle\delta_h({\bf k})\delta_m({\bf k})^*\rangle}{\langle\vert\delta_m({\bf k})\vert^2\rangle}
\end{equation}
and the stochastic term $\varepsilon({\bf k})$ is assumed to be uncorrelated with the dark matter field (i.e., $\langle\varepsilon({\bf k})\delta_m({\bf k})^*\rangle = 0$). Note that $b(k)\delta_m({\bf k})$ is completely correlated with the dark matter field. Hence, the measured halo spectrum $\hat{P}_{hh}(k)$ can be decomposed into two terms:
\begin{equation}\label{eq:13}
\hat{P}_{hh}(k) = P_{hh}(k) + \sigma^2(k)
\end{equation}
with
\begin{equation}\label{eq:14}
P_{hh}(k) = b^2(k)P_{mm}(k),
\end{equation}
where $\sigma^2(k) = \langle\vert\varepsilon({\bf k})\vert^2\rangle$ is the stochastic noise and $P_{mm}(k) = \langle\vert\delta_m({\bf k})\vert^2\rangle$ is the dark matter spectrum. This decomposition makes the stochastic noise $\sigma^2(k)$ contain all sources of stochasticity between halos and dark matter (\citealt{2009PhRvL.103i1303S}). For uniformly weighted halo field, the stochastic noise $\sigma^2(k)$ is usually modeled as the Poisson shot noise, which is given by the inverse of the halo number density $1/\bar{n}_{h}$. If each halo is weighted by $w_i$, the expected shot noise is generalized to be
\begin{equation}\label{eq:15}
\sigma_{exp}^2(k) = \dfrac{V\Sigma_{i=1}^Nw_i^2}{(\Sigma_{i=1}^Nw_i)^2},
\end{equation}
where $V$ is the volume of simulation box and $N$ is the number of halos.

Halo stochasticity can lead to a lack of coherence between halo field and dark matter field, and thus is usually described by the \emph{cross-correlation coefficient}:
\begin{equation}\label{eq:16}
r_{hm}(k)\equiv\frac{P_{hm}(k)}{\sqrt{\hat{P}_{hh}(k)P_{mm}(k)}} = (1+\dfrac{\sigma^2(k)}{b^2(k)P_{mm}(k)})^{-1/2},
\end{equation}
where $p_{hm}(k)$ is the cross-power spectrum of the two fields. Here, we utilized equation (\ref{eq:13}, \ref{eq:14}) for deriving the right hand side of equation (\ref{eq:16}), which explicitly indicates that the existence of stochastic noise $\sigma^2(k)$ will make $r_{hm}(k)$ deviate from unity and the extent of deviation also depends on the bias $b(k)$. If there is no stochasticity (i.e., $r_{hm}(k) = 1$), it means that the spatial distribution of dark matter can be derived from that of halos once the bias $b(k)$ is known (\citealt{2009MNRAS.396.1610B}). In particular, the equation (\ref{eq:16}) can be further rewritten as
\begin{equation}\label{eq:17}
\frac{\sigma^2(k)}{P_{hh}(k)}=\frac{1-r_{hm}^2(k)}{r_{hm}^2(k)},
\end{equation}
which is called as the \emph{inverse signal-to-noise ratio} (hereafter N/S ratio) of the halo spectrum and is also used to describe the halo stochasticity\footnote{Other statistical descriptions of halo stochasticity also can be found in literatures, e.g., $S(k)\equiv\sqrt{2(1-r_{hm}(k))}$ (\citealt{2004MNRAS.355..129S}; \citealt{2009MNRAS.396.1610B}) and $E(k)\equiv\sqrt{1-r_{hm}^2(k)}$ (\citealt{2011MNRAS.412..995C}). Nevertheless, they all depend only on the cross-correlation coefficient between halos and dark matter, including the equation (\ref{eq:17}).} (cf. \citealt{2010PhRvD..82d3515H}).

\section{Results}
\label{sec:results}

\subsection{The halo clustering bias}
\label{sec:clustering bias}

The halo clustering bias is scale dependent, especially on small scales. A significant scale dependence of the bias can be treated as a sign of non-negligible deviations from zero for the higher-order bias parameters in equation (\ref{eq:6}). Nevertheless, on large scales ($k \lesssim 0.1$ $h$Mpc$^{-1}$), it is expected to be a constant, which corresponds to the linear bias parameter in equation (\ref{eq:6}). This constant offset in the large-scale clustering amplitude relative to dark matter can be corrected to reconstruct the dark matter power spectrum (\citealt{1999MNRAS.308..119S}), while due to the bias scale-dependence and other uncertainties the small-scale galaxy/halo clustering information is usually discarded. Moreover, the more massive halos should be more biased, because they have relevant higher clustering.

In figure \ref{fig:bias}, we plot the biases of the different weighted halo fields by using equation (\ref{eq:12}). For the uniform weighting case, we see that with the increase of halo number density, the bias on small scales (i.e., $k \gtrsim 0.1$ $h$Mpc$^{-1}$) becomes more scale dependent. We checked that this is a combined result of various bias behaviors of halo populations with different masses\footnote{For checking this, we also measured the biases of different halo mass bins. Specifically, we split our largest sample into 10 mass bins with equal logarithmic mass intervals, given that halo abundances drop sharply with the increase of mass. Indeed, we find that the biases of different halo populations have different trends on small scales, i.e., it tends to rise for high-mass halos and tends to fall for low-mass halos.}, which is also applicable to other weighting cases.

In our study, we take the average of the bias on scales $0.025$ $h$Mpc$^{-1} \lesssim k \lesssim 0.099$ $h$Mpc$^{-1}$ (corresponding to the shaded regions in figure \ref{fig:bias}) to serve as the linear bias parameter (cf. table \ref{tab:info}). It is worth noting that for the two halo samples with higher halo number densities, the optimal weighting makes their biases more scale independent relative to the other two weighting schemes. This may provide a more reliable way to determine dark matter spectrum directly from the optimally weighted halo spectrum by only using the linear bias model, which thus helps to extract precise information from large-scale structure surveys (e.g., the primordial non-Gaussianity, signatures of massive neutrinos, etc.) and to minimize the systematic shifts in the BAO position relative to the dark matter (cf. \citealt{2008MNRAS.383..755A}; \citealt{2008PhRvD..77d3525S}; \citealt{2008arXiv0802.2416Z}). In addition, for this study, bias being more scale independent also helps to adequately maximize the performance of bias tracer reconstruction by only considering the linear bias parameter of equation (\ref{eq:6}) (i.e., setting higher-order bias parameters to be zero).

\subsection{The stochastic noises of different weighted halo fields}
\label{sec:stochastic noise}

In figure \ref{fig:shot_noise}, we show the \emph{expected} shot noises of different weighted halo fields (i.e., the solid lines), obtained by using equation (\ref{eq:15}). And, if halos are randomly distributed in space, for any weighting scheme, the corresponding bias should be zero (as the corresponding random halo field has no correlation with the dark matter field, i.e. $\langle\delta_h({\bf k})\delta_m({\bf k})^*\rangle = 0$, cf. equation (\ref{eq:12})), thus, the corresponding halo power spectrum is completely contributed by the stochastic noise (cf. Eq.(\ref{eq:13}) and Eq.(\ref{eq:14})). Here, we call this noise as the \emph{simulated} shot noise, which should in principle be equal to its Poisson expectation. Indeed, as shown in figure \ref{fig:shot_noise}, the simulated shot noises (i.e., the asterisk points) are well matched with their expectations, which could serve as a nicety test of our numerical implementations.

Actually, due to halo exclusion\footnote{Halo exclusion refers to the fact that two halos cannot be too close to each other arbitrarily, corresponding to the halo two-point correlation function $\xi_{hh}(r)$ approaching $-1$ at the halo separation $r$ less than the average diameter of halos.} and clustering, which violate the Poisson assumption of placing point particles randomly in space, the halo stochastic noise turns out to be non-Poissonian (see \citealt{2013PhRvD..88h3507B}; \citealt{2017MNRAS.470.2566P}; \citealt{2017PhRvD..96h3528G} and Refs. therein). Here, we also call the halo stochastic noise as the \emph{measured} shot noise, which is obtained by using equation (\ref{eq:13}). The halo exclusion and non-linear clustering can lead to the measured shot noise sub- and super-Poissonian, respectively. And, the amplitudes of these deviations from Poisson expectation also depend on the bias parameters and the mass ranges (\citealt{2013PhRvD..88h3507B}). Specifically, with larger exclusion scales and higher linear bias parameter, the high-mass halos are most affected by the exclusion effect, while the non-linear clustering effect is most important for low-mass halos, where the value of the second-order bias parameter is non-zero and the exclusion effect is small (see \citealt{2013PhRvD..88h3507B}; \citealt{2017MNRAS.470.2566P} for more detailed discussions). As shown in figure \ref{fig:shot_noise}, for the uniform weighting case, the non-Poissonian behaviors of the measured shot noises are actually caused by the two competing effects.

For the mass and optimal weighting cases, the measured shot noise amplitudes are significantly below the Poissonian predictions (cf. figure \ref{fig:shot_noise}), which have been demonstrated in \citealt{2009PhRvL.103i1303S} and \citealt{2010PhRvD..82d3515H}. This phenomenology indicates that the mass-dependent weightings can bring some extra information for considerably suppressing shot noise component. We argue that these information should originate from the environment dependence of halos to meet the requirements of local mass and momentum conservation (\citealt{2009PhRvL.103i1303S}), considering that the masses of halos have tight relations with the halo local environments (e.g., the local halo number density within some distance) (\citealt{2012MNRAS.419.2133H}; \citealt{2015MNRAS.451.4266Z}). We can erase these information by shuffling the masses and positions of halos separately and randomly recombining them. After this procedure, the measured shot noises of the shuffled halos therefore have similar non-Poisson behaviors as the uniform weighting case, where these noises are dubbed as the \emph{shuffled} shot noise in this paper (cf. appendix \ref{sec:appendix A}).

\subsection{The correlation between halos and dark matter}
\label{sec:cross-correlation}

In figure \ref{fig:stochasticity}, we present the cross-correlation coefficients between different weighted halo fields and dark matter field, which describe the similarities with the dark matter field. In addition, we also show the N/S ratios of different weighted halo power spectra (cf. equation (\ref{eq:17})) in figure \ref{fig:stochasticity}, which helps amplify the differences at large scales where the cross-correlation coefficients are very close to unity. Benefiting from suppression of shot noise and boosting of clustering bias (cf. figure \ref{fig:bias} and equation (\ref{eq:16})), the mass and optimally weighted halo fields have markedly stronger correlations with the dark matter field (cf. \citealt{2009PhRvL.103i1303S}; \citealt{2010PhRvD..82d3515H}), which means that these weighted halo fields retain more information of dark matter distribution.

Generally, all above presented results suggest that mass-dependent weightings could be useful to improve the precision of cosmological parameter estimation and to reduce undesirable systematics of BAO measurements by using biased tracers. In particular, mass-dependent weightings could potentially lead to dramatic improvements in the efficiency of the biased tracer reconstruction, which we will investigate in the following sections.

\subsection{Direct reconstruction without preprocessing bias}
\label{sec:direct reconstruction}

To quantify the initial information successfully recovered by the reconstruction algorithm (or called as reconstruction performance), we calculate the cross-correlation coefficients between the simulation’s initial condition and the different weighted halo fields before and after reconstruction, i.e., $r_{XI}(k)\equiv\frac{P_{XI}(k)}{\sqrt{P_{XX}(k)P_{II}(k)}}$, where "X" and "I" stand for the field of pre- or post-reconstruction and the initial condition respectively (cf. figure \ref{fig:reconstruction_without_bias}). As reference, the results for the dark matter field case are also shown. Compared to the underlying dark matter, the discrete and biased tracers suffer from various additional complications (e.g., the sparseness and bias etc.), which will undoubtedly degrade the reconstruction performance in the practical applications. Therefore, in the following tests, the cross-correlation coefficient between the reconstructed dark matter field and the initial condition could serve as an upper bound on the initial information, which can be recovered from a halo field.

We first investigate the consequences of direct reconstruction without any preprocessing of clustering bias (i.e., setting $b_1=1$ and $\gamma_2=b_2=0$ respectively), which can help understand the bias effects on the reconstruction performance. The results are presented in figure \ref{fig:reconstruction_without_bias}, where the upper and bottom panels respectively show the power spectra and the cross-correlation coefficients with the initial condition for different fields of pre- and post-reconstruction. Additionally, in the top panels, the power spectrum of the initial condition is also plotted as a reference.

We see that the process of reconstruction mainly modifies the power spectra on relatively small scales (i.e., $k > 0.1$ $h$Mpc$^{-1}$) and the power spectra of pre- and post-reconstruction have similar shapes on sufficiently large scales (also see \citealt{2017ApJ...847..110Y}). In general, the reconstruction improves the correlation coefficients depending on the considered halo number density (cf. \citealt{2017ApJ...847..110Y}; \citealt{2019MNRAS.483.5267B}). We find that the reconstruction performances for different weighting schemes obey $r_{UI}(k) < r_{MI}(k) < r_{OI}(k)$ ("U", "M" and "O" denote uniform, mass and optimal weighting respectively) on very large scales, while the relationship among them becomes uncertain on relatively small scales which should be attributed to the wrong assumption of bias parameters (cf. \citealt{2019ApJ...870..116W}; \citealt{2019MNRAS.483.5267B}). In particular, for the optimal weighting, we show an excellent result for the halo sample with the highest number density, as the corresponding bias is quite close to unity and also very scale independent (cf. figure \ref{fig:bias} and table \ref{tab:info}).

\subsection{The reconstruction considering linear bias}
\label{sec:considering linear bias}

The bias between tracers and underlying dark matter is an important limiting factor for the efficiency of reconstruction. Inappropriate treatment of bias in reconstruction will lead to errors in the estimation of the displacement field and thus significantly degrade the recovery of the initial matter distribution (\citealt{2019ApJ...870..116W}). \citealt{2019MNRAS.483.5267B} found that the performance of biased tracer reconstruction is most sensitive to the linear bias parameter and is only marginally affected by higher-order bias parameters in their tests. The main purpose of this work is to validate the improvements on the performance of biased tracer reconstruction by considering halo mass information. In view of the quasi-scale-independent features of the biases under mass-dependent weightings (mostly for the scenarios of higher halo number densities, cf. figure \ref{fig:bias}), we will only consider $b_1$ (i.e., setting  $\gamma_2=b_2=0$) in the reconstruction procedure for our performance tests.

Before proceeding to test the performance quantitatively, we first show a visual comparison of different fields after reconstruction. The two-dimensional slices of projected overdensities are presented in figure \ref{fig:density_field_after_reconstruction}, where for the reconstructed halo fields we only show the case of $\bar{n}_h = 2.77\times10^{-2}$ ($h^{-1}$Mpc)$^{-3}$ to quote our results\footnote{Other halo number density cases can be found in the appendix \ref{sec:appendix B}.}. We see that the reconstructed halo field under uniform weighting seems to be somewhat noisy, while for the other two mass-dependent weighting cases, the reconstructed halo fields are relatively smoother, which could be treated as a benefit brought by the information of halo masses. Visually, we find that the reconstruction results for the dark matter and the optimally weighted halos are quite similar (cf. figure \ref{fig:density_field_after_reconstruction}), showing a great potential of optimal weighting for highly improving the efficiency of biased tracer reconstruction.

The quantitative results are presented in figure \ref{fig:reconstruction_with_bias}. We also plot $S_{XI}(k) \equiv \sqrt{2(1 - r_{XI}(k))}$ to amplify the differences at large scales, where $r_{XI}(k)$ is very close to unity. We find that the linear debiasing can largely improve the performances for the mass and optimal weighting cases, while it's a little complicated in the uniform weighting case. Specifically, for the uniform weighting case, the bias tends to be more scale dependent (i.e., the bias changes more significantly with increase in $k$) on small scales with the increase of halo number density (cf. the yellow curves in figure \ref{fig:bias}). For the scenarios with higher halo number densities, this implies that a simple linear debiasing is inadequate and the more sophisticated non-linear debiasing schemes are needed for fully maximizing the performance. Therefore, for $\bar{n}_h = 2.77\times10^{-2}$ and $2.77\times10^{-3}$ ($h^{-1}$Mpc)$^{-3}$ cases, the $r_{UI}(k)$s' drop more steeply with the increase of $k$ on small scales, compared to the direct reconstruction (cf. the solid yellow lines in the bottom panels of figure \ref{fig:reconstruction_without_bias} and middle panels of figure \ref{fig:reconstruction_with_bias}).

Nevertheless, under the uniform weighting, the linear debiasing indeed boosts $r_{UI}(k)$ on relatively large scales ($k < 0.3$ $h$Mpc$^{-1}$) for $\bar{n}_h = 2.77\times10^{-2}$ ($h^{-1}$Mpc)$^{-3}$, while for $2.77\times10^{-3}$ ($h^{-1}$Mpc)$^{-3}$ the $r_{UI}(k)$ basically has no improvements on scales of $k > 0.2$ $h$Mpc$^{-1}$ after debiasing $b_1$, which should be due to the fact that the $b_1$ is close to unity (here $b_1 = 0.94$; cf. figure \ref{fig:bias} and table \ref{tab:info}). And, for the massive halo sample with the lowest number density, we find that the linear debiasing is quite important to improve the performance, since the bias deviates from unity and is also very scale independent (cf. figure \ref{fig:bias}). These results suggest the importance of linear debiasing in the reconstruction-related cosmological studies, where the large-scale information is more concerned. For example, the power spectrum from a typical galaxy survey basically contains BAO signals extending to the scale of $k = 0.3\sim0.4$ $h$Mpc$^{-1}$, signals on scales smaller than which are not detectable due to the poor signal-to-noise ratio, thus the recovery of information larger than this scale is enough to restore the BAO signals.

\begin{figure*}
\begin{minipage}{\textwidth}
\raisebox{0.1cm}{\includegraphics[width=1\columnwidth]{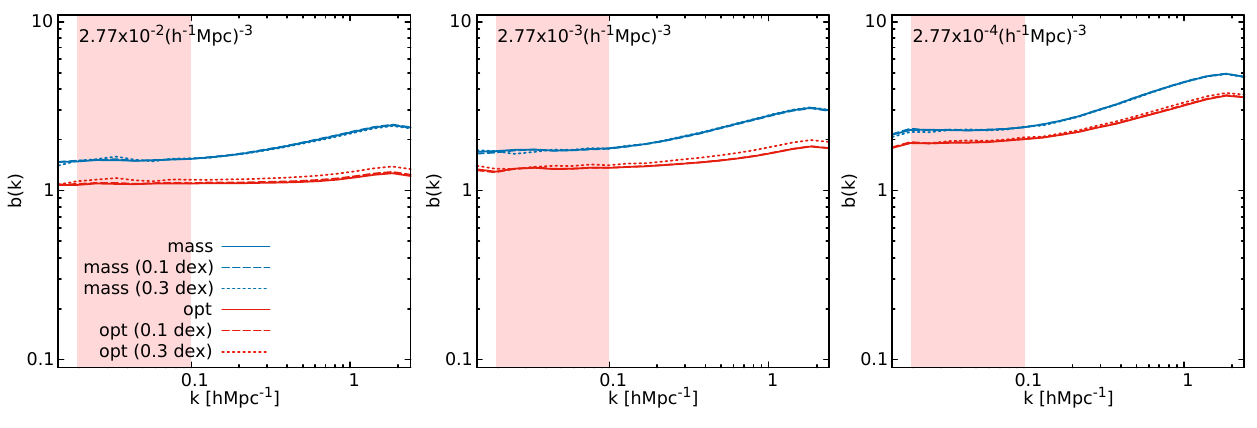}}
\caption{The clustering biases of the mass and optimally weighted halo fields with various halo mass scatters. The dashed lines and dotted lines correspond to the mass scatters of 0.1 dex and 0.3 dex, respectively (the same below). For comparison, the clustering biases of mass and optimally weighted halo fields without mass scatters (solid lines) are also presented.}
\label{fig:bias_with_mass_scatter}
\end{minipage}
\end{figure*}

\begin{figure*}
\begin{minipage}{\textwidth}
\raisebox{0.1cm}{\includegraphics[width=1\columnwidth]{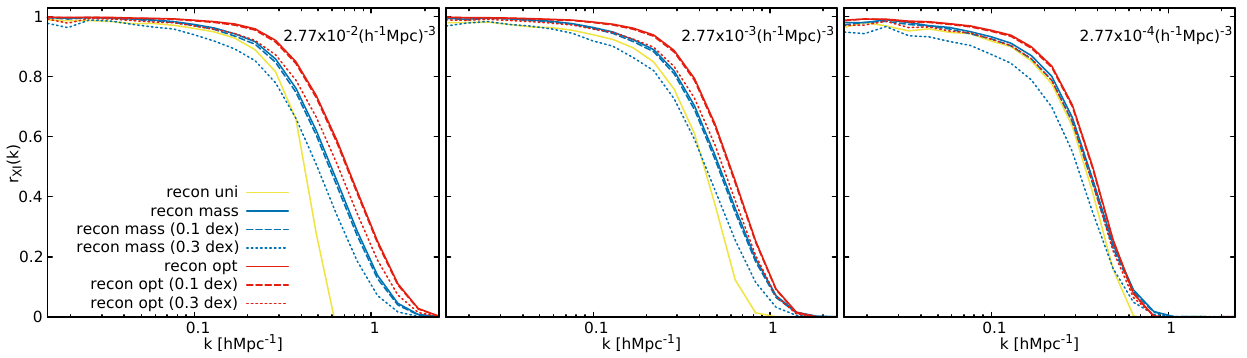}}
\caption{The cross-correlation coefficients with initial condition for the reconstructed mass and optimally weighted halo fields with various halo mass scatters. For comparison, we also plot the cross-correlation coefficients with initial condition for various reconstructed halo fields without mass scatters (solid lines).}
\label{fig:reconstruction_with_mass_scatter}
\end{minipage}
\end{figure*}

On the whole, as expected, the performance indeed obeys $r_{UI}(k) < r_{MI}(k) < r_{OI}(k)$ after linear debiasing in the procedure of reconstruction, consistent with the conjecture in the former sections. This due to fact that halo fields under the mass-dependent weighting schemes reserve more information of dark matter distribution and have lower shot noises. On the other hand, the validity of this reconstruction algorithm is based on the assumption of mass conservation in LSS evolution (cf. equation \ref{eq:1}), and mass-dependent weightings can be treated as suitable implementations of the idea to enforce the local mass and momentum conservation, which reasonably improve the efficiency of the algorithm. This conclusion should also apply to the isobaric reconstruction (\citealt{2017PhRvD..96l3502Z}) and other improved algorithms (\citealt{2017PhRvD..96b3505S}; \citealt{2018MNRAS.478.1866H}; \citealt{2020arXiv201209074L}), which also use mass conservation to relate the density fields in Lagrangian and Eulerian spaces. In particular, the optimal weighting has much greater advantages to improve the reconstruction performance relative to the usually used uniform weighting scheme, and can even work perfectly for the halo sample with the highest number density. This should make the optimal weighting quite interesting for the on-going and future galaxy surveys (e.g., DESI and LSST) to measure the BAO characteristic scale and other cosmological observables, which suffer from the nonlinearities of LSS.

\subsection{The effects of mass scatters on reconstruction performances}
\label{sec: mass scatters}

The key information required to achieve these gains are the halo masses, which, however, are not directly observable in reality. Nevertheless, in principle, they are achievable via the relations between observables [e.g., galaxy luminosity (\citealt{2004MNRAS.353..189V}), X-ray luminosity, galaxy richness, weak lensing shear, velocity dispersion or the thermal Sunyaev-Zel’dovich (SZ) effect (cf. \citealt{2011ARA&A..49..409A}) etc.] and halo masses. A better understanding of these relations should be essential and critical to minimize the the inaccuracies of inferred masses and maximize the gains of biased tracer reconstruction.

The SZ effect and X-ray properties yield very tight correlations with the halo masses with a scatter of about 0.1 dex, whereas these observables can only be used to probe the masses of massive systems (e.g., galaxy clusters), while the optical estimators can be used to infer relatively lower halo masses with a scatter of about 0.18 dex (cf. \citealt{2019MNRAS.489.2439H}; \citealt{2020MNRAS.493..337B} and Refs. therein). Alternatively, galaxy grouping methods (\citealt{2005MNRAS.356.1293Y}), which group galaxies residing in the same halo, provide a direct way of studying the galaxy–halo connection to estimate halo masses. These techniques can reliably estimate halo mass over a wide range of masses, even for a poor system including isolated galaxy in a small halo. By using a halo-based group finder, \citealt{2007ApJ...671..153Y} demonstrated that more than $90\%$ of true halos can be successfully identified in mock catalogues and the scatter between estimated and true halo masses is about 0.3 dex. This group finder was also improved for applications in low-redshift galaxy surveys, showing a lower mass scatter of about 0.2 dex (\citealt{2017MNRAS.470.2982L}). Recently, \citealt{2020arXiv200605426W} developed a machine-learning-based group finder, which was applied to high-redshift incomplete spectroscopic data, estimating halo masses with scatters smaller than 0.25 dex.

In order to mimic the observational uncertainties on the estimated halo masses, we artificially add two constant mass scatters (i.e., 0.1 dex and 0.3 dex for optimistic and pessimistic scenarios respectively) to our halo samples, and then repeat the analyses for all halo number density cases shown in figure \ref{fig:bias_with_mass_scatter} and figure \ref{fig:reconstruction_with_mass_scatter}. For the two mass-dependent weightings, we find that the effects of a scatter of 0.1 dex on the reconstruction performances are basically negligible and a scatter of 0.3 dex can result in noticeable degradations of the performances. For the scatter of 0.3 dex, the benefits from mass weighting are completely lost, while the benefits from optimal weighting still remain substantial in the higher number density cases, which again demonstrates the interests of optimal weighing in the applications of on-going and future galaxy surveys. Here, note that the optimal weighting adopts the same $M_{0}$s' (see section \ref{sec:halo fields}) as before. We also see that the biases are hardly affected by the mass uncertainties (cf. figure \ref{fig:bias_with_mass_scatter}), which indicates that the performance degradations are purely caused by the boosts of shot noises induced by the mass scatters (cf. \citealt{2009PhRvL.103i1303S}).

\section{Summary and Conclusion}
\label{sec:conclusions}

In realistic scenarios, the efficiency of reconstruction is prone to be hampered by various complications (e.g., galaxy bias, shot noise, RSDs and survey boundary, etc.). In principle, the problem could be alleviated by considering extra information (e.g., the radial velocities for reducing boundary effects (\citealt{2020MNRAS.494.4244Z}) and the local environments for sharpening the BAO peak (\citealt{2015PhRvD..92h3523A}; \citealt{2019MNRAS.482..578B}) etc.).

The biased tracer reconstruction method (\citealt{2019MNRAS.483.5267B}) adopted in this work uses mass conservation to relate the density fields in Lagrangian and Eulerian spaces. When the method is applied to a realistic tracer field, the reconstruction performance will be degraded by the loss of information of underlying total matter. The basic idea for improving the reconstruction performance should be to apply reconstruction to a tracer field with more matter distribution information, which can be achieved from some weighting schemes. Our work corroborated this idea by using halo mass-dependent weighting schemes, since halo masses are natural information to enforce the local mass and momentum conservation.

For our study, we performed careful analyses on three halo samples with various number densities corresponding to different targets in various galaxy surveys. The halo fields are created under different halo weighting schemes, i.e., uniform, mass and optimal weightings (because the mass and optimal weightings are both dependent on the information of halo masses, we collectively call them as mass-dependent weightings). We summarize our work and give discussions as follows:

\begin{itemize}
      \item   Before our performance tests, we first investigated the clustering biases and stochasticity properties of different weighted halo fields, demonstrating how mass information suppresses the shot noise component and tightens the cross-correlation between the halo field and the dark matter. We argue that the information for suppressing the halo field’s shot noise should originate from the environmental dependence of halo mass (cf. section \ref{sec:stochastic noise} and appendix \ref{sec:appendix A}). For our halo samples with $\bar{n}_h=2.77\times 10^{-2}$ $(h^{-1}\mathrm{Mpc})^{-3}$ and $\bar{n}_h=2.77\times 10^{-3}$ $(h^{-1}\mathrm{Mpc})^{-3}$, we find that halo clustering biases are more scale independent under mass-dependent weightings, compared with the uniform weighting case. This finding is interesting since bias being more scale independent will help to adequately maximize the performance of the bias tracer reconstruction by only performing linear debiasing in the process of reconstruction and to minimize the systematic shifts in the BAO position relative to the dark matter. For this reason, we only considered the clustering bias up to linear order in our performance tests.

     \item   For performance tests, we then performed the biased tracer reconstruction method on the different weighted halo fields. We investigated and discussed in detail how linear debiasing improves the reconstruction performance. After linear debiasing, we both visually and quantitatively compared the qualities of the reconstructed initial matter fields, which are obtained under different weighting schemes. We showed that the reconstruction performance can be substantially enhanced when the reconstruction method is applied to the mass and optimally weighted halo fields, compared to the uniform weighting scenario. As the halo number density increases, the gains achieved from mass information can be more significant. In particular, we showed a compelling result that the reconstruction performance from the optimally weighted halo field with $\bar{n}_h=2.77\times 10^{-2}$ $(h^{-1}\mathrm{Mpc})^{-3}$ can almost be comparable to that from the dark matter field, which implies that the initial information can be greatly recovered from the dense data [e.g., the DESI Bright Galaxy Survey (BGS) sample (\citealt{2016arXiv161100036D})]. Nevertheless, the reconstruction performance also depends on the accuracy of inferred halo masses in the observations, thus the impacts of realistic halo mass scatters on our results were also investigated in this work. A better understanding of the relation between observables and underlying halos in the galaxy-halo connection and observable-mass relation studies (see \citealt{2018ARA&A..56..435W}; \citealt{2020MNRAS.493..337B} and Refs. therein) should be essential and critical to decreasing the mass scatters and consequently to maximizing the gains of biased tracer reconstruction.
\end{itemize}

Based on our findings, we can safely conclude that the halo masses are critical information for highly improving the performance of biased tracer reconstruction. We note that this conclusion should also hold true for other reconstruction methods, as the effectiveness of almost all reconstruction methods relies on the information of total matter distribution.  

Our work is particularly relevant to the context of recovering the BAO signal, the characteristic scale of which is renowned as a standard ruler to measure the expansion history of the Universe, and are therefore of critical importance for future BAO measurements. We leave the quantification of the gains of the BAO reconstruction benefiting from the halo mass information to future investigations. Moreover, since various other cosmological measurements (e.g.,RSDs (\citealt{2012PhRvD..86j3513H}), primordial non-Gaussianities (\citealt{2020arXiv200708472D}) and even neutrino signatures (\citealt{2020PhRvD.101f3515L})) are also usually limited by strong data nonlinearities, we expect our work to be quite useful for significantly improving the scientific returns of current and future galaxy surveys.

\section*{Acknowledgements}
We would like to thank Pengjie Zhang, Haojie Xu, Zhejie Ding, Ji Yao and Hong-Ming Zhu for helpful discussions and several useful suggestions and the anonymous referee for the helpful comments. Y.L. would also like to thank Xiaohu Yang and Qingyang Li for useful communications about galaxy group finder. This work was supported by the National Key Basic Research and Development Program of China (No. 2018YFA0404504), the National Science Foundation of China (grants No. 11773048, 11621303, 11890691) and the "111" Project of the Ministry of Education under grant No. B20019. BL is supported by the European Research Council (ERC) through a Starting Grant ERC-StG-716532-PUNCA, and by the UK Science and Technology Facility Council through grants ST/T000244/1 and ST/P000541/1.

\appendix

\section{The shuffled shot noises}
\label{sec:appendix A}

\begin{figure*}
 \centering
  \includegraphics[width=1\linewidth]{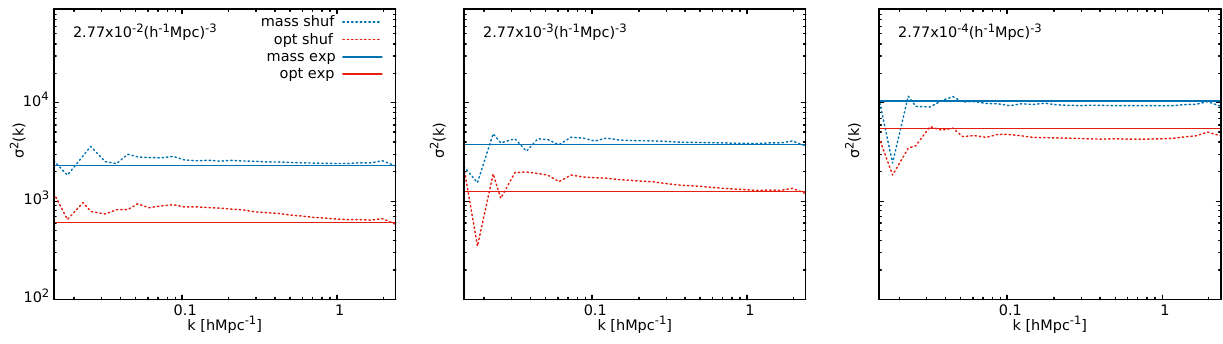}
  \caption{The shuffled shot noises under mass and optimal weighting schemes (dotted lines). For comparison, the expected shot noises under mass and optimal weighting schemes (solid lines) are also presented.}
 \label{fig:shuffled_shot_noise}
\end{figure*}

In this appendix, we present the shuffled shot noises under mass-dependent weightings, which are mentioned in section \ref{sec:stochastic noise}. The results are shown in figure \ref{fig:shuffled_shot_noise}.

\section{Different weighted halo fields before and after reconstruction}
\label{sec:appendix B}

\begin{figure*}
\begin{minipage}{\textwidth}
\center
\raisebox{0.0cm}{\includegraphics[width=0.97\columnwidth]{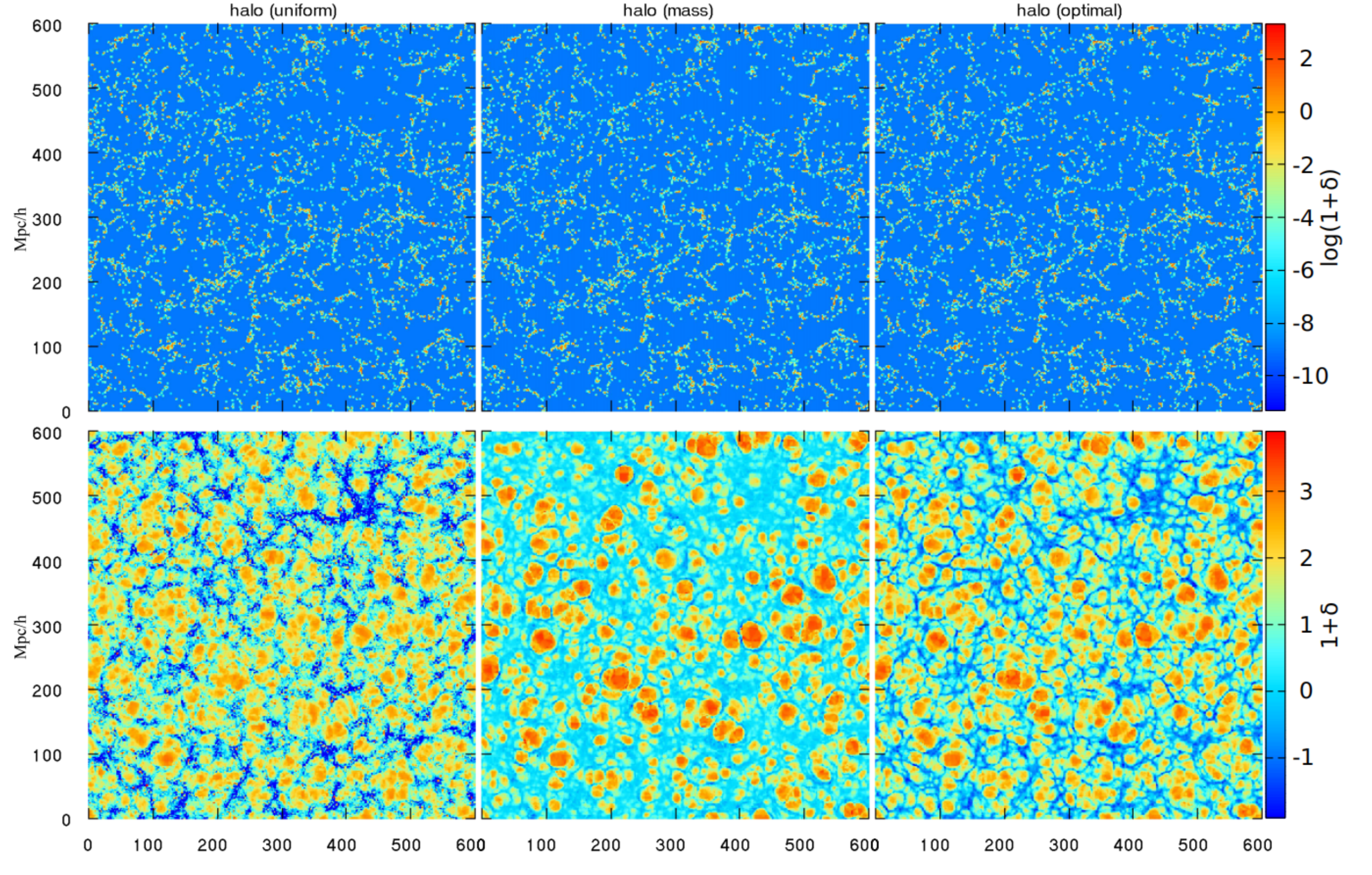}}
\caption{The two-dimensional slices of different weighted halo fields before (upper panels) and after (bottom panels) reconstruction. The halo fields are produced by the halo sample with number density of  $2.77\times10^{-3}$ $(h^{-1}{\rm Mpc})^{-3}$. Each slice shows the same region (with the same projection depth) as the corresponding slice in figure \ref{fig:density_field_before_reconstruction} and figure \ref{fig:density_field_after_reconstruction}.}
\label{fig:density_field_middle}
\end{minipage}
\end{figure*}

\begin{figure*}
\begin{minipage}{\textwidth}
\center
\raisebox{0.0cm}{\includegraphics[width=0.97\columnwidth]{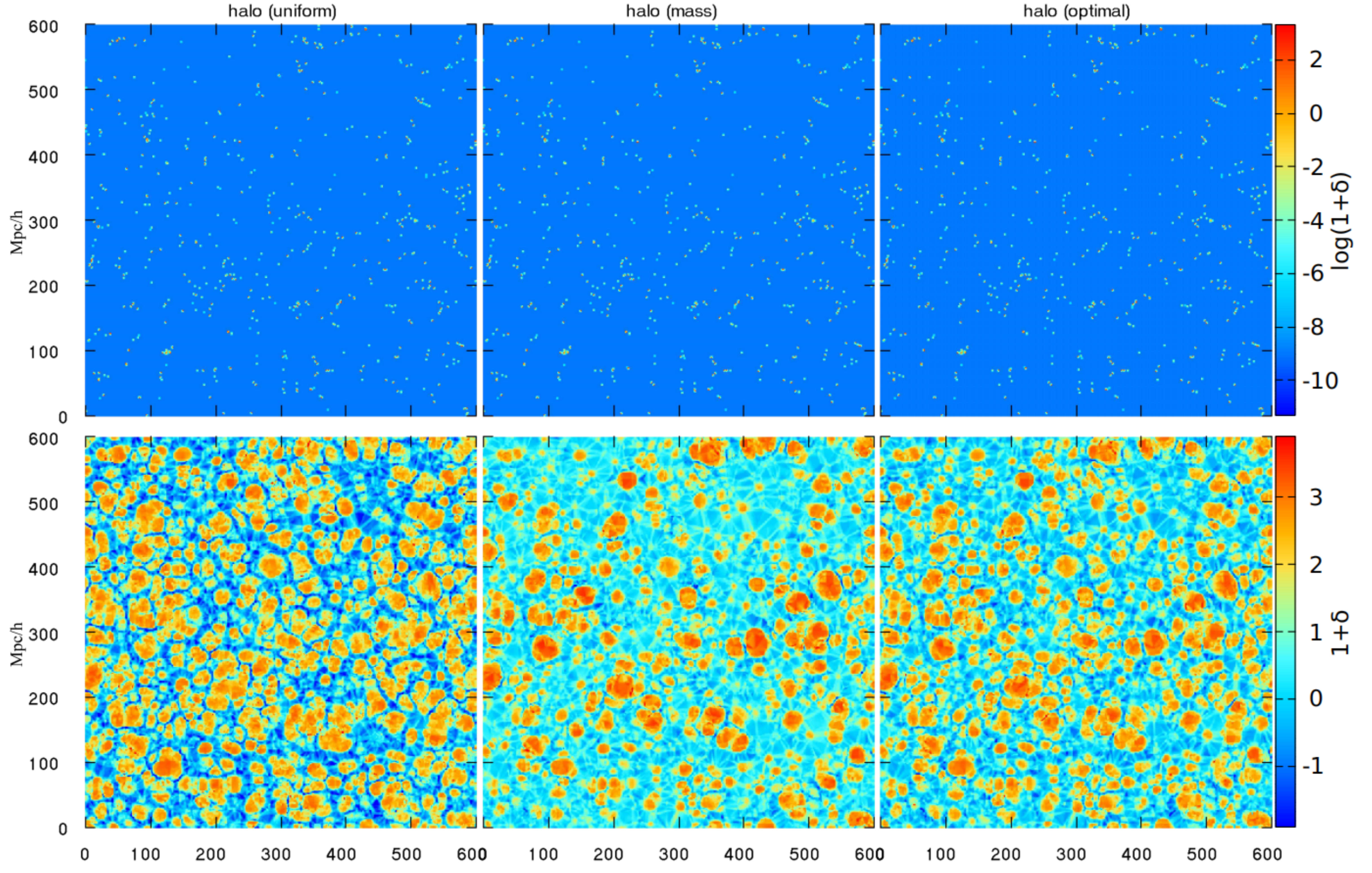}}
\caption{Same as figure \ref{fig:density_field_middle}, but for the halo sample with number density of  $2.77\times10^{-4}$ $(h^{-1}{\rm Mpc})^{-3}$.}
\label{fig:density_field_bottom}
\end{minipage}
\end{figure*}

In this appendix, we present the two-dimensional slices of different weighted halo fields with number densities of $2.77\times10^{-3}$ and $2.77\times10^{-4}$ $(h^{-1}{\rm Mpc})^{-3}$ before and after reconstruction. The results are shown in figure \ref{fig:density_field_middle} and figure \ref{fig:density_field_bottom}.

\bibliographystyle{aasjournal}  %mnras_mwilliams}
\bibliography{ref}

% required because of bug in MN2e style file
% throws away figs otherwise
\clearpage

\end{document}